# Probing Charge Transfer Dynamics in a Single Iron Tetraphenylporphyrin Dyad Adsorbed on an Insulating Surface

*Pablo Ramos[1], Marc Mankarious[1,2], Michele Pavanello[1], Damien Riedel[2*]*

[1] Department of Chemistry, Rutgers University, Newark, New Jersey 07102, USA

[2] Institut des Sciences Moléculaires d'Orsay (ISMO), CNRS, Univ. Paris Sud, Université Paris-Saclay, F-91405 Orsay, France

**Abstract:**

**Although the dynamics of charge transfer (CT) processes can be probed with ultimate lifetime resolution, the helplessness to control CT at the nanoscale constitutes one of the most important road-blocks to revealing some of its deep fundamental aspects. In this work, we present an investigation of CT dynamics in a single iron tetraphenylporphyrin (Fe-TPP) donor/acceptor dyad adsorbed on a $CaF_2$/Si(100) insulating surface. The tip of a scanning tunneling microscope (STM) is used to create local ionic states in one fragment of the dyad. The CT process is monitored by imaging subsequent changes in the neighbor acceptor molecule and its efficiency is mapped revealing the influence of the initial excited state in the donor molecule. In validation of the experiments, simulations based on density functional theory show that holes have a higher donor-acceptor CT rate compared to electrons and highlight a noticeable initial state dependence on the CT process. We leverage the unprecedented spatial resolution achieved in our experiments to show that the CT process in the dyad is governed via molecule-molecule coherent tunneling with negligible surface-mediated character.**

**Introduction**

Charge transfer (CT) processes are the cornerstone of a plethora of physical and chemical phenomena including electronic transport[1], solar cell design[2], organic emitting devices[3], spintronics[4], and even photosynthesis in biosystems[5]. CT processes occurs intrinsically at the nanoscale across this wide variety of disciplines[6] because it is a quantum process that is influenced by various parameters such as the molecular symmetry, the vibrational modes or the ergodicity of the studied systems[7]. For these reasons, CT processes have been intensively studied in gas phase or in solution where various parameters can be controlled to tune the CT rate such as the nature of the solvent or the presence of functionalized substituents[8]. In this context, several experimental techniques offer the ability to probe the CT rate, ranging from adsorption/emission spectroscopy[9] to more sophisticated optical pump-probe experiments[10]. However, difficulties arise because these techniques mainly report on an average of various spatial configurations restraining the capacity to simultaneously control morphology, energetics and energy dissipation pathways. On the molecular modeling side, CT processes involve electronic states possessing drastically different spatial shape from the electronic ground state. This leads to commonplace theories such as the time-dependent density functional theory (DFT) to falter[11]. The combination of such complications have somehow hampered progresses in the fundamental understanding of CT in molecular assemblies[12,13].

In order to bring new insights in the investigation of CT processes, it is necessary to modify the currently adopted paradigms. One enticing alternative is to study CT at the nanoscale in a well-defined and controlled environment. Self-assembled molecules on metallic surfaces[14], vertical electronic transport with STM[15] or planar broken junctions[16] are systems and techniques offering interesting and rich physics in this context but do not warrant electronic decoupling of the molecule with the surface and are recognized to often provide ill-defined conformational environments of the entire studied system. Low temperature STM can tackle these deficiencies because it offers the ability to control the environment down to the Angstrom-level and can be employed as an initiator of CT by generating local ions on targeted regions of a single molecules[17,4,18,19]. Specifically, when molecules are adsorbed on a surface, the influence of symmetry breaking resulting of the formation of nonstationary localized charge states can be characterized with specific molecular reactions (movement, switch, charge storage, luminescence)[20, 21].

Recent investigations dealing with individual heterodimers located on an metal/insulating surface have been performed to investigate energy transfer processes when probed via harvesting of the electronically induced luminescence signal[22,23]. Expectedly, the described energy transfer process is shown to be insensitive to the tip localization over the donor molecule since couplings for energy transfer are of longer range than for CT processes and follows very different pathways as they are influenced by specific factors[24,25]. So far, controlled CT processes have never been investigated in molecular dyads at the nanoscale when the initial nonstationary state of CT is prepared precisely at specific locations inside one of the molecular fragments[26].

In this work, we focus our study on a molecular dyad made of two unbonded iron-5,10,15,20-tertaphenyl-21H,23H-porphyrin (Fe-TPP) molecules that plays the role of donor-acceptor homodimer when physisorbed on a thin insulating layer (Fig. 1a). Porphyrins and in particular metalated porphyrins represent a particularly important class of CT systems as they are involved in a large number of biological or physical processes[27,28,29,30]. While tunnel electrons from STM are traditionally used to image surfaces and adsorbates, they also allow the formation of very local vibrational excitation or transient ions via electronic activation to investigate various physical or chemical phenomena such as molecular dissociation[31], switching[32] or luminescence emission[33]. In this work, we study electron and hole transfer in the dyad with a low temperature STM (9 K) by triggering the CT through the creation of an ionic state in the donor molecule leading to a charge transfer to the acceptor[34]. The lateral movement of the acceptor along the surface is witnessed and correlated to the exchange of charge between monomers. Here, the robustness of a very well documented excitation process to control molecular manipulation[21,17,19,20,38,39,41,42] represents a corner stone to develop a complete novel approach to study charge transfer process at the nanoscale. Our investigations show that the transfer of holes is favored in the molecular dyad compared to the transfer of electrons. We also report that the hole transfer rate depends on the location where the ionized state (i.e. the initial state) is generated in the donor monomer. Thus, we have concentrated our investigation on the spatial dependence of the CT event, an elusive aspect of CT[35], especially in systems where the environment is unable to polarize and stabilize the initial and final states involved in a CT process[24,25]. By investigating the influence of the CT process on the excitation location in the donor fragment, we show that the ensuing CT process arises from a coherent tunneling effect where possible surface mediated channels appears to be negligible. This new method of investigating CT at the nanoscale can be employed to specifically study bridge-mediated CT (e.g., heterodimers featuring various molecular bridges involving covalent or non-covalent bonding), opening a new door to studying CT at the nanoscale in extremely well-defined and controlled environments.

**Results and discussion:**

In the gas phase, Fe-TPP exhibits $D_{4h}$ symmetry having a well characterized structure and a ground electronic state that is summarized in Fig. 1b[36]. In this work, several Fe-TPP molecules are gently adsorbed on the cooled insulating $CaF_2$ stripes on Si(100) as described previously[18]. The large scale STM picture (Fig. 1c) depicts Fe-TPP molecules adsorbed on top of the insulating stripes which arrange into two main adsorption configurations named CL and CR (see smaller scale STM pictures shown in Figs. 1d and 1e)[37]. The CL configuration (Fig. 1d) shows an anticlockwise rotation of its skeleton of 23° from the insulating stripe direction. This implies that the upper right and lower left phenyl groups are rotated to be almost coplanar to the porphyrin macrocycle (see white arrows in the right panel of Fig. 1d). The other two opposite phenyls are less impacted by the morphology of the insulating stripes because they are located on top of the groove that separate two insulating stripes. This gives them more freedom to adjust the dihedral angle $\theta$ between the phenyl and the porphyrin. The CR configuration (Fig. 1e) corresponds to the mirror symmetry of the CL conformation with a clockwise rotation (same angle as CL) of the molecule compared to the stripe leading to a similar rotation of the upper left and lower right phenyl groups (see white arrows in the right panel of Fig 1e). For sake of clarity, we report the optimized molecular geometries superimposed to the STM images in the right panels in Figs. 1d and 1e. The size of the bright protrusion considered as a Fe-TPP molecule is coherent with the one of the gas phase molecule (Fig. 1b) as shown in the apparent height profile provided in the Supplementary Fig. S1. Owing to the relatively large surface energy gap of the insulating layer, the presented STM topographies involve deeper occupied molecular orbitals, as in this case, the frontier HOMO and LUMO orbitals are not reachable via STM imaging.

To investigate CT in molecular homodimers, it is necessary to characterize the electronic induced motion of a single molecule as it will be used as a probe of the CT process in the dyad. The study of a single Fe-TPP molecule movement is performed by exciting electronically each of the two CL and CR conformations at eight different points as indicated in Figs. 2a and 2c. Specifically, points 1 to 4 correspond to excitation positions where the STM tip is on top of the phenyl groups of the molecule whereas points 5 to 8 designate excitations located on the pyrrole groups of the porphyrin macrocycle. Due to mirror symmetry between the CR and CL molecular conformations, the excitation positions are numbered in a way that allow a direct point-to-point motion yield comparison. The excitation is carried out with two different biases, probing unoccupied orbitals (positive bias + 2.5 V) or occupied orbitals (negative bias - 2.5 V). The excitation procedure is as follows: (i) The STM tip apex is placed at one of the excitation location defined above. (ii) The feedback loop of the STM is switched off and the bias is adjusted to a value of - 2.5 V or + 2.5 V. (iii) during the excitation time, the tunnel current intensity is recorded and a variation of tunnel current is detected when the molecule has moved[38,39]. Subsequently to this excitation process, the feedback loop is turned on again and the STM recovers its previous scanning parameters. (iv) the molecule is subsequently imaged to observe any conformational changes for which a molecule having a CL (or CR) conformation can switch to a CR (or CL) configuration after the electronic excitation time leading to two possible cases. The first corresponds to a CL → $CR_{up}$ (CR → $CL_{up}$) transition. The second case is when the molecule slides and rotates upward along the insulating stripe via a CL → $CR_{down}$ (CR → $CL_{down}$) movement (*i.e.*, when the molecule slips and rotates upwards (downward) along the same stripe). The insulating stripe serves then as a track on which the molecule can move as previously reported[18]. Occasionally, when the excited molecule hops and lands on a neighboring stripes (different from the initial one) or when the molecule moves several steps upward or downward, the results of the manipulations are disregarded from the presented statistical analysis. To check that the excitation-induced conformational change is similar from one molecule to the other, we have also performed sets of measurements on various different molecules of each configuration located at various places on the surface as well as on different terraces. For each excitation points, the efficiency to induce the molecular motion is inferred from these measurements and expressed as a quantum yield $Y = (e/I_{exc.} \times T_{exc.})$ where $e$ is the electron charge, $I_{exc.}$ the tunnel current intensity during the excitation and $T_{exc.}$, the time needed to induce the conformation change[39,40,41]. This method is further explained in the Supplementary Fig. S2.

The measured yield of the CL → CR movement when a single Fe-TPP molecule is initially excited in a CL conformation (Fig. 2a) is presented in Fig. 2b. The figure shows that both biases (-2.5 V and +2.5 V) are efficient for moving the molecule upward or downward when the excitation is applied locally at points 1-4 (*i.e.*, at the phenyl groups). The yield strongly decreases when the excitations are applied in points 5-8 (*i.e.*, at the pyrroles groups) of the molecule. A careful look in Fig. 2b unveils interesting trends. The upward movement is favored for both biases when the excitation is located at points 2 and 4 while the downward movement is more efficient when the excitation is placed at points 1 or 3 (see the insert in Fig. 2b). We also note that excitation at negative bias leads to slightly higher yields than positive bias.

The excitation of a Fe-TPP initially in a CR conformation (Fig. 2c) can be switched through a $CL_{up}$ or $CL_{down}$ movement independently of the excitation bias (Fig. 2d). However, in order to move upward the corresponding Fe-TPP molecule (i.e., in the $CL_{up}$ conformation), it is required to apply the excitation at points 1 or 3. The downward movement ($CL_{down}$) is instead favored with an excitation at the locations 2 or 4. We stress here that the excitation positions located at point 5 to 8 do not result in significant molecular motions yields for both CL and CR conformations. The apparently random yields measured at these positions may arise from spurious electronic excitations in the neighborhood of the considered positions.

The STM-induced molecular movement of the single Fe-TPP molecule on the surface is a one-electron process that forms a transient cation (-2.5 V) or anion (2.5 V)[39,42]. During the transient lifetime of the ionic fragment (for example a cation), the equilibrium geometry of the ionized molecule differs from the neutral species because of the image charge in the substrate as generally described in Antoniewicz processes[43]. This effect leads generally to an excess of kinetic energy once the molecule is neutralized[43]. When the bias changes (*i.e.*, from -2.5 V to +2.5 V), the molecular picture is qualitatively the same, although the molecular ion spends some time in an anionic state. During the lifetime of the ionic state (and thus during its quantum evolution) the molecular motion is initially triggered via a repulsive electrostatic effect due the electronic charge of the STM tip being of the same sign of the ion[42]. The molecular movement then continues via the gained kinetic energy after the molecule becomes neutral as further detailed in the Supplementary Fig. S3.

Our experimental data indicate a major effect – i.e., the motion occurs preferentially when the excitation is initially applied on one of the phenyl groups of the Fe-TPP molecule. When the phenyl groups are excited, the motion of the molecule is favored because of their various degrees of freedom compared to the porphyrin macrocycle. Hence, the localization of the excitation allows us to select a specific region of the potential energy surface (PES) of the molecule in which the transient local ion is formed, involving a superposition of non-stationary excited states in the molecule. Because of possible favorable symmetries of the involved wavefunctions, the electronic interaction between the phenyl rings and the porphyrin macrocycle will significantly depends on the dihedral angle between these molecular groups[17,44,45]. Taking this electronic interconnection into account, it is possible to exploit the described molecular motion induced by a transient ionic state as a probe of the CT process in a molecular homodimer.

The following task is to assemble two Fe-TPP molecules in a specific conformation to form a dyad. For simplicity, we consider hereafter only homodimers made of two CL molecules due to mirror symmetry with the CR-CR dyad. In the dyad, the upper molecule acts as the donor and the lower molecule as the acceptor (Fig. 3a). This arrangement allows us to use the STM tip-induced ionization to study CT from donor to acceptor. The overall process exploits the created ion to trigger either a hole or an electron transfer to the neighbor molecule. The neutralization of the second molecule induces its movement along the stripe and constitutes a signature of the CT process that occurs in the homodimer.

To gain a comprehensive picture of the CT process, we position the STM tip at four different locations (1 to 4) of the acceptor molecule, each corresponding to a pyrrole groups (Fig. 3a). We point out that we do not excite the phenyl groups of the donor to avoid probable motion of the excited molecule and hence favor the CT process to the acceptor molecule. Therefore, from the initial CL-CL dyad configuration, only the CL → CR$_{down}$ molecular conformation change is considered (right panel in Fig. 3a) as a probe of the CT process. For each of the four selected excitation positions and biases, a probability to trigger the CT process in the dyad is deduced and the lower molecule of the dimer is replaced with the STM tip to reform the initial CL-CL dyad conformation. The measured quantitative quantum yield as a signature of the CT process is then extracted from this probability (see the method section). The ensuing measured CT yields are presented in Fig. 3b for the two considered biases -2.5 *V* and 2.5 V.

Our results show that when the excitation is applied at positive voltage, the yield to trigger the CT process in the homodimer is very low. However, for negative bias (-2.5 V), the CT quantum yield is significantly improved for the four selected positions and is found to strongly depend on the chosen excitation location. This first observation indicates that the hole transfer is favored over electron transfer. To provide a better understanding of these observations, we plot the variation of the CT efficiency as a function of the distance that separates the excitation location to the center of the neighbor molecule (Fig. 3c). Each excitation position is thus defined with a distance $d_1$ to $d_4$ and the resulting curve for both biases are presented in Fig. 3d. Since the two molecular fragments of the homodimer are not chemically bonded to each other, it is interesting to investigate which type of CT process occur in the dyad. Generally, CT involving tunneling processes are favored in unbounded systems compared to resonant CT which typically requires specific spatial distribution of density of states[11]. The trends in Fig. 3d for hole transfer (-2.5 V) show that the CT process triggered in the Fe-TPP homodimer does not only depend on the chosen excitation location but also on the distances between the excitation positions and the acceptor molecule. Our results reveal that the location at which the CT process is initiated strongly influence the CT rate and hence the lifetime of the excited ionic state. In other words, there is a significant dependence of the hole transfer yields on the preparation of the initial state of the CT that is not observed for the excitation at positive bias, i.e. for electron transfer (Fig. 3d).

Further characterization of the CT processes is provided by simulations based on the density-functional theory (DFT) which aim at corroborating the following observations: **(1)** the CT is observed to be more efficient for holes than for electrons; **(2)** the CT efficiency depends on the position at which the hole is generated in the donor molecule. That is, there is a clear and noticeable dependence of the CT rate on the preparation of the initial nonstationary charged-localized state.

Our first investigation has the goal to define the relevant geometrical configuration of the Fe-TTP molecules in the dyad on the insulating surface stripes. From the experimental observation, upon adsorption, the molecule is rotated by an angle of $\gamma = 23°$ around the axis that crosses perpendicularly the central metal atom of the molecule. Additionally, due to steric hindrance with the surface, two phenyl groups are rotated of an angle $\theta$, placing them in a plane almost similar to the one of the porphyrin macrocycle (Fig. 4a). Taking $\theta$ as an important parameter for the CT rate, we computed the electronic coupling $V_{ij}$ of the CT (see method) from donor to acceptor for three different values of $\theta$: 10°, 45° and 90°. The dyad is made with one molecule as in the gas phase while the second one sees its structure modified according to the values of $\theta$ as explained in Fig. 4a. The value $\theta = 90°$ corresponds to a dihedral angle where the phenyl plane is perpendicular to that of the porphyrin macrocycle. Our calculations show that for $\gamma \sim 20°$, the electronic coupling is optimal when $\theta = 10°$ which is consistent with our experimental observations. By plotting the spatial distribution of the frontier orbitals HOMO and LUMO (Fig. 4b) one can see that a large overlap of the HOMO orbitals over the porphyrin, unlike LUMO, can clearly enhance the values of $V_{ij}$. Further delocalization of the HOMO is driven when the phenyl group's rotation angle $\theta$ is such that it brings the HOMO orbital to hybridize over the rotated phenyl groups (Fig. 4b)[46,47]. A detailed description of these simulations is given in the Supplementary Fig. S4.

Our DFT simulations can also quantitatively examine the CT rate dependence on the position of the initial charged-localized state as well as the difference between hole and electron transfer processes. To reproduce the experimental excitation locations, we have defined four atomic groups, i.e., the atoms constituting the pyrrole moieties of the donor molecule (Fig. 5a). The CT rate $\Gamma(E)$ is evaluated by the Fermi Golden Rule expression. Our model addresses the rate of neutralization of the donor molecule after it has been ionized and is given by:

$$\Gamma(E) = \frac{2\pi}{\hbar} \sum_{i \in \text{initial}} P_I(\varepsilon_i) f_I(\varepsilon_i - \mu_I) \delta(E - \varepsilon_i) \sum_{j \in \text{final}} \theta_{ij} f_F(\varepsilon_j - \mu_F) |V_{ij}|^2, \quad (1)$$

with:

$$\theta_{ij} = \begin{cases} \Theta(\varepsilon_j - \varepsilon_i) & \rightarrow \text{for hole transfer,} \\ \Theta(\varepsilon_j - \varepsilon_i)\Theta(\varepsilon_j - \varepsilon_{LUMO}) & \rightarrow \text{for electron transfer.} \end{cases} \quad (2)$$

Here, E is the tunneling energy of the electron or hole leaving the molecule, $P_I$ is the partial density of states (PDOS) of the upper (donor) excited molecule calculated at each pyrrole. The $f_{I/F}$ are Fermi-Dirac functions pertaining the initial/final states and $\Theta(\varepsilon)$ is the Heaviside step function at the energy $\varepsilon$ that ensures thermodynamic irreversibility. $V_{ij}$ is the electronic coupling (defined in the methods section) connecting the electronic states of donor and acceptor, $\mu_{I/F}$ are the chemical potential of each fragment (Fermi Energy, $E_F$), and $\varepsilon_{i/j}$ are the values of the energy levels and $\delta(\varepsilon)$ is the Dirac delta function at the energy $\varepsilon$.

Equation (1) shows that the calculated CT rate probes the reachable CT pathways between the two molecules in the dyad when one of them (donor) is initially ionized (with a loss of an electron) at the state with energy $\varepsilon_i$. Subsequently, the charge (hole) is transferred to the states of the acceptor molecule lying in the energy window $\varepsilon_j - \varepsilon_i$. The localized STM induced excited ionic state is simulated by weighing the transfer rate by the partial density of electronic states PDOS (indicated by $P_I$ in equation (1)) at a chosen pyrrole group. Similarly, the contribution of the final state is given by $\frac{2\pi}{\hbar}|V_{ij}|^2$ which is the Fermi golden rule rate for the $i \rightarrow j$ transition. Therefore, the CT rate $\Gamma(E)$ depends on the excitation position in relation to the symmetry of the molecule and its ensuing spatial DOS distribution at each considered orbital. In our simulations we assume that the spatial distribution of the involved orbitals does not change drastically between the neutral and the ionic states. The error introduced by this approximation on the calculated couplings in porphyrin systems was previously assessed to be below 20 % and therefore acceptable for qualitative comparison with experimental findings[48]. Additional information about the rate described in Eq. (1) can be found in the methods section.

The values of $\Gamma(E)$ are calculated for a fixed distance (16 Å) between the donor and acceptor Fe-TPP molecules (Fig. 5a). The molecules are rotated by an angle of 23° to the left to mimic the CL-CL conformation as observed experimentally. The influence of the surface on the Fe-TPP morphology is taken into account by locking a set of two given dihedral angles between the two previously defined phenyl groups and the porphyrin (Fig. 4a) and then monitoring their effect on the CT rate. We have considered the computation of two particular cases for which the upper (donor) molecule is fixed either with $\theta = 45°$ or $\theta = 10°$. In both cases, the bottom (acceptor) molecule is kept in the same configuration with $\theta = 10°$ accordingly to the experimental observation. The variations of $\Gamma(E)$ as a function of the energy $E - E_F$ are presented in Figs. 5b and 5c for the electron transfer and in Figs. 5d and 5e for hole transfer for the two particular values of $\theta$ described above. For electron transfer (Figs. 5b, 5c), we notice a significant CT rate peak centered at 2.3 eV with a maximum value of $\Gamma(E)$ in the range 0.3-0.4 x $10^6$ s$^{-1}$. We also observe that the variation of $\Gamma(E)$ intensity for electron transfer does not strongly vary for both value of $\theta$ as a function of the excitation position. For hole transfer (Figs. 5d, 5e), the major $\Gamma(E)$ peaks are mainly centered at -1.8 eV. The intensity of the $\Gamma(E)$ peaks for hole transfer in Figs. 5d and 5e are more than two orders of magnitude higher than for the electron transfer (i.e. $0.1 - 0.45$ x $10^8$ s$^{-1}$). A careful look at the maximum intensity of these peaks at -1.8 eV in Figs. 5d, 5e clarifies how the adsorption conformation of the Fe-TPP molecule strongly influences the hole CT rate as $\Gamma(E)$ is more than twice larger for $\theta = 10°$ than for $\theta = 45°$. Plotting the maximum of the peak intensities of $\Gamma(E)$ at -1.8 eV and 2.5 eV as a function of the distance of the excitation point to the acceptor ($d_1$ to $d_4$) reveals a set of curves with a parabolic-like shape for the hole transfer whereas $\Gamma(E)$ remains rather flat as a function of the distance for the electron transfer. These data can be compared with the experimental curves shown in Fig. 3d. Comparing electron and hole transfer rates allows us to conclusively determine that the hole transfer occurs with higher rates than electron transfer. If we concentrated now on the calculated hole transfer rate (Fig. 5f), we can see that for the three shorter distances ($d_2$, $d_3$, $d_4$), the calculated variation of $\Gamma(E)$ for $\theta = 10°$ reproduces the trends of the variations of the experimentally measured CT yields (Fig. 3d). However, the calculated hole CT rate for the larger distance $d_1$ increases whereas it stays relatively steady for the experimental values. This difference suggests that the hole CT rate depends not only on the excitation position but also on the distance between the initial excited state and the acceptor molecule.

Inspection of the involved molecular electronic states reveals that different orbital symmetries can affect the CT rates. Hence, to understand the role of the molecular conformation in relation to its ensuing symmetry on the variations of the calculated CT rate, it is important to draw a detailed picture of the DOS spatial distribution for each molecular orbitals. Such distribution is presented in Fig. 6 for the first five LUMOs and the fourteen HOMOs. For biases spreading in the range 0 to 2.5 V, the number of probed LUMOs orbitals is weak and their DOS are mainly spanning within the iron atom and the porphyrin macrocycle. In the range 0 to -2.5V, a much larger number of orbitals are involved. In particular, there are clear asymmetric distributions of DOS in the HOMO-11 and HOMO-12 where one of the pyrroles groups shows almost no density of state (red arrows in the lower part of Fig. 6). Other similar asymmetric DOS distributions at the pyrroles groups are also observed (*i.e.*, HOMO-5 and HOMO-6) for which the DOS is spreading through the porphyrin macrocycle and on the rotated phenyl groups at $\theta = 10°$ (HOMO-5). A comparison with the DOS distribution of a single Fe-TPP molecule without surface perturbation (*i.e.*, no rotation along $\theta$) indicates that the asymmetric distribution of DOS in the molecular orbitals mainly arise from the angle variation between the rotated pairs of phenyl groups in the Fe-TPP (see supplementary Fig. S5).

**Discussion**

Considering the Antoniewicz-like process that describes the molecular motion of a single Fe-TPP molecule (Fig. S3), we coherently use the same model to depict the CT mechanism occurring in the Fe-TPP dyad in relation to our experimental and theoretical observations. When the STM tip creates a hole at one of the selected pyrrole group of the donor molecule (Fe-TPP$_1$ in Fig. 7a), the resulting PES of the pyrrole group (Pyr$^+$) reaches a higher energy curve that matches the energy of a specific orbital of the acceptor molecule (e.g. Porph + Aryl$^N$ at Fe-TPP$_2$ in Fig. 7a). For a given excited pyrrole, the ensuing quantum state of the local excited ionic PES varies in relation to the initial local DOS distribution differences between pyrroles at a given energy. Therefore, while the local ionization potential can be considered as identical at each pyrroles, the electronic coupling between each fragments combined with the Frank-Condon factors in the dyad can thus lead to a specific hole transfer dynamics from the donor to the acceptor molecules during the life time of the excited ionic state[17]. The second step (Fig. 7b) of the CT process describes the evolution of the excited ionic state of the acceptor fragment (Porph + Aryl$^+$) following the loss of one electron. Here we illustrate the most favorable case (θ = 10°) for which the porphyrin macrocycle and the aryl groups involved in the final molecular motion of the acceptor fragment are coupled (see the double well potential PES in Fig. 7a and 7b). Simultaneously, the electron incoming from the acceptor fragment that neutralizes the donor moiety provides additional kinetic energy (*ke*) which is lost within the donor molecule/substrate interactions without inducing any molecular motion. Subsequently, as a third and final step, the neutralization of the acceptor fragment (Fe-TPP$_2$ in Fig. 7c) allows the molecule to gain a kinetic energy *ke'* that leads to the excitation of vibrational modes smearing over the molecule and in particular at the aryl groups of the second fragment, which rules its motion along the insulating stripe.

Indeed, as observed experimentally, the movement of the Fe-TTP molecule is enhanced when the electronic excitation of the molecule is applied on one of the aryl group of the porphyrin. Therefore, the movement of the second fragment is optimized if the majority of the gained kinetic energy, *ke*, is coupled to the rotational and vibrational population of the phenyl groups located at position 1 or 3 on the CL conformation (Fig. 2b). A careful look in Fig. 6 show that there are a few particular molecular orbitals that can answer these criteria and thus favor the ensuing CT process: they lie at higher energy than the excited orbital of the donor fragment and their amplitude spreads both on the porphyrin cycle and on one or two of the considered aryl groups. This is particularly the case for the HOMO-5 orbital for which the rotated phenyls at position 2 and 3 allow to spread the DOS over the active part of the molecule. A similar structure can also be observed at the HOMO orbital where the excitation of the phenyl at position 3 can induce the observed molecular motion (see red arrow in Fig. 6).

At this point of our investigation, it is possible to address the question of whether the CT process occurs in the molecular dyad as a tunneling process or a sequential hopping CT process that may involve several hole transfer steps. More precisely, it would be interesting to learn from our investigations if the hole created at one precise location at the pyrrole PES of the Fe-TPP is rapidly transferred to the HOMO-5 orbital of the acceptor fragment or if the CT process occurs following an entire delocalization of the hole through the porphyrin macrocycle PES of the donor fragment following the excitation. In the latter case, the measured yield is expected to show weak reliance upon the considered distances and would involve a rapid relaxation to the frontier occupied orbital HOMO of the donor. In this configuration analogous to the Kasha rule[49], the measured CT yield would not show variations when the excitation energy of the tunnel electron varies. This process is related to the charge transfer time. CT occurring through frontier orbitals of the dyad can be estimated to take place in the picosecond time scale[50]. However, another relevant information relies on the comparison of the life time $\tau$ of the CT process when it involves each possible occupied orbital in the donor fragment, with a CT process life time $\tau_{HOMO}$ that will only occur from the frontier (HOMO) orbital of the donor. The inset in Fig. 6 shows the variations of the ratio $\tau/\tau_{HOMO}$ as a function of the relative energy $E-E_F^{Tip}$ and indicates that the CT in the dyad is three order of magnitude faster if it occurs from deeper occupied orbitals of energy ranging from -2.5 to -1.8 eV whereas the CT process slows down if higher-energy orbitals (i.e. -1.7 to HOMO) are involved. This implies femtosecond timescale for hole transfer from orbitals deeper than the HOMOs. Another analogous CT time estimation arising from the measured CT yields is provided in the supplementary Note N5.

To further highlight the CT dynamics, we have fitted the experimental data presented in Fig. 3d for the hole transfer yield with an exponential decay function of the form $\Gamma(d) = \gamma_0 + \gamma_1 \exp(-d/\beta)$, where $\gamma_0$ and $\gamma_1$ are proportionality and offset factors, respectively, and $\beta$ is a falloff parameter as described in the McConnell relation that describes CT processes involving tunnel electrons[51]. A satisfactory set of parameters allow us to plot the resulting fitting function in Fig. 3d (red dashed curve) with a value of $\beta = 2.2 \pm 0.5$ Å$^{-1}$. Our results reveal that the hole CT process observed in the dyad rather arises from a coherent tunneling effect initiated at each of the pyrrole groups of the excited porphyrin since the value of $\beta$ is relatively large compared to what can be observed in other biosystems with stronger donor-acceptor electronic communication[52]. Hence, for the observed CT process in the FeTPP dyad, a relatively short life time of the excited ionic state is expected, preventing that the created hole at deeper orbitals (i.e. HOMO-11) relax rapidly over the entire donor molecule to a frontier orbital. Although this relaxation channel cannot be completely excluded as a competing relaxation pathway, it is not a determining factor in the observed CT process. The excess of energy that is provided by the electronic excitation at a slightly higher energy of the ionization threshold of the molecule allows to prepare a transient cationic state in an excited specific roto-vibrational state that matches the ones of the acceptor molecule. Such a resonance condition favors a CT process due to high Franck-Condon factors[7,17], hence displaying anti kasha behavior. This is not surprising as anti-Kasha process is generally favored in systems with a low number of collisions and where the on-site energy dissipation is slow[49]. The trend of our findings is also consistent with our experimental conditions performed at low temperature (9 K) and thus rather exclude the possibility of a purely thermally-activated processes.

At this stage, we found it essential to study a possible influence of the electrostatic field variations at the acceptor molecule when the donor molecule is excited at the positions 1-4 with the STM tip. The estimation given in the supplementary Fig. S6 shows that the maximum variation of the electrostatic field reaches ~3% in the worst case (tip radius = 10 nm and micro-tip radius = 1 nm, between $p_1$ and $p_4$). Such a small variation cannot explain our experimental findings for which the measured CT yields and the calculated rates $\Gamma(E)$ differences worth ~30% between the two nearest positions $p_3$ and $p_4$. Additionally, it is important to notice that the absolute electrostatic field variations is the same when the STM tip shifts from position p1 to p4 while the bias is changed from -2.5V to +2.5 V. Yet, the measured CT yield exhibit sharp differences of almost one order of magnitude between these two biases (Fig. 5f). Therefore, it is conclusive to say that our investigation of the CT process in a FeTPP dyad is not perturbed by the electrostatic field variations in the STM junction. Others interactions induced by the electrostatic field present in the STM junction on the molecular dyad are related to the energy shifts of the molecular orbitals levels. In a double junction, this shift can reach ~ 0.5 eV[42]. This effect will not impede the describe CT process but may rather results in picking an orbital near the calculated Γ(E) resonance (~1.8 eV) below the Fermi energy of the surface.

We can also discuss the role of the surface and in particular if the CT process in the dyad is related to a surface mediated process. This can be explored via the dI/dV curves acquired on the molecule and on the CaF$_2$ insulating layer. The dI/dV curves traduce the presence of DOS at a given energy and shows that near the excitation bias (-2.5 V), a peak of density of states located at a slightly lower energy (-2.75 eV) in the silicon surface could be involved in a surface mediated process (see supplementary Fig. S7 ). Here, the DOS band spreading from -1 to -2 V in the dI/dV curve (Fig. S7) arises from the presence of a group of deeper molecular orbitals as calculated in Fig. 6 (i.e. HOMO-5 to HOMO-14). To distinguish between a (coherent) tunneling CT process in the dyad and one involving a surface mediated effect, we have performed electronic excitation directly on the insulating surface at two different distances as being one or two times the distance *d* that separate the two molecular fragments in the dyad. The ensuing results indicate that the yield to move the Fe-TPP at a given distance (*d* or *2d*) is more than three order of magnitude lower than when the excitation is applied directly on the molecule. Surprisingly, electronic excitation applied on the surface at -2.5 V are almost one order of magnitude lower than the one applied at 2.5 V. These results rule out the possibility of having a CT process in the dyad involving surface state of the insulating layer (see supplementary figure S7). Interestingly, the fact that the molecule can be perturbed via the direct excitation of the insulating layer may involve a completely different process and in particular excitation via the propagation of exciton or a guided light within the insulating layer. This will probably conduct to very fascinating future studies with similar molecular devices at the nanoscale.

**Conclusion**

Our work show that it is possible to manipulate, generate and investigate few-molecule assemblies as model structures for studying charge transfer processes at the nanoscale on surfaces. By using the tunnel electrons of the STM tip, it is possible to create local transient ions that trigger various CT processes via hole or electron transfer. Here, we choose a homomolecular dyad formed by two Fe-TPP molecules adsorbed on a thin insulating surface of monolayer $CaF_2$/Si(100) resulting in a typical donor-acceptor system. The very high spatial precision of the STM allows us to select specific initial PES as transient ionic states of the donor molecule. Our results show that this method can optimize the CT rate when it is initiated at various positions in the donor molecule. It also demonstrates that the precise molecular conformations of the donor and acceptor molecules as well as their relative position have significant influences on the CT rate. Furthermore, our investigations run at low temperature (9K) reveal that the ensuing CT process arises mainly from a tunneling effect whose dynamics appears to have negligible interactions with the surface and through electrostatic field variations in the STM junction. This excludes a major influence of surface mediated or hopping CT processes. Thus, our method for investigating CT at the nanoscale is versatile and can be extended to study the influence of various molecular conformation of covalent or noncovalent molecular assemblies including bridged complexes. Combined with other techniques (such as the luminescence analysis emitted in the dyad) our approach opens the door to a large variety of new investigations in relation to biological systems, the improvement of solar cells and the understanding of light emitting and charge storage devices for which the interface interactions between the active molecular media with different types of electrodes still lack a deep understanding and control. In addition, we have also showed that it is possible to generate and handle local ionic states which display typical characteristics of charge separated states. Thus, our work shows that CT events can be inspected in real time at the atomic and molecular length scales.

## Acknowledgments:

This material is based upon work supported by the National Science Foundation under Grant No. OISE-1404739, and by the U.S. Department of Energy, Office of Science, Office of Basic Energy Sciences under Award Number DE-SC0018343. DR acknowledges CNRS for financial support through the *programme international de collaboration scientifique* (PICS) THEBES, contract No. PICS07244 and the LUMAT federation for their financial supports.

# Methods:
## Experimental methods:

The experiments are performed with a low temperature (9K) scanning tunneling microscope (STM) working in ultra-high vacuum (UHV). The surfaces are prepared from highly doped (n-type, As doped, $\rho = 5$ m$\Omega$.cm) Si(100) samples. The bare silicon surface is reconstructed in a c(4x2) structure as explained in several previous works via multiple annealing cycles. To minimize surface defects, the base pressure in the UHV is kept under $4 \times 10^{-11}$ torr during this process. A thin $CaF_2$ layer is then grown while keeping the silicon surface at ~ 1050 K. The evaporation of the $CaF_2$ molecules is performed via a second effusion cell heated at ~ 1350 K with an exposure of 1.3 monolayer. The obtained epitaxial surface is then cooled down. Sequentially, the Fe-TPP molecules are evaporated on the silicon surface by heating a Knudsen cell at ~550 K (i.e. below their dissociation temperature ~ 673 K). Through this process, the surface is kept at low temperature (12 K) via a liquid helium cooling of the sample holder to warrant a soft landing on the substrate and reduce irreversible surface-molecule interactions. In addition, a low evaporation rate is chosen to reduce the formation of molecule clusters on the surface. These parameters are adjusted with a quartz balance to obtain an homogenous molecular coverage > 0.1 ML. The Fe-TPP molecule is chosen as a model molecule for its relatively low ionization potential and its interest in the transport of apical ligands in Heme. The formation of the noncovalent dyad is performed in-situ by molecular manipulation. Due to the use of a low temperature STM (9 K), the reduced lateral drift during the excitation process or the dI/dV measurements is very low (~ 0.02 Å) and thus warrant the precision and repeatability of our measurements. The excitation method of the molecular dyad is slightly different to the one used to study the single Fe-TPP movement, since, in the dyad the molecule that moves due to a CT process is not underneath the STM tip and thus can hardly be recorded in the tunnel current trace. Hence, for the dyad, we use a pre-defined excitation duration $T_{exc.}$ estimated to provide a probability to induce the CT lower than 1 (i.e. the mean activation time $t_0$ is lower than $T_{exc.}$). The tunnel current (i.e. the tip height) during the excitation procedure is then slightly adjusted with this criteria and to check that the process involve only one electron[53]. Because the ensuing probability can change with the tunnel current intensity, it is more accurate to express the molecular change probability via a quantum yield. Hence, for a measured probability of success $p_s$ to induce the CT in the dyad (which is the ratio between the number of successful excitation and the total number of excitation) at an excitation current $I_{exc.}$, a quantitative estimation of the yield of the CT process (per tunnel electron) is traduced by the value $Y = e/(I_{exc.} \times t_0)$, where $I_{exc.}$ is the excitation current, $e$ the charge of the electron and where $t_0 = -T_{exc}/ln(1-p_s)$ is extracted from a binomial law[39,40].

The dI/dV measurements have been performed with a double lock-in amplifier that modulate the voltage bias at a frequency $f = 847$ Hz with an amplitude of ~10 mV. The dI/dV measurements are repeated several times at the same positions at different tip heights and averaged over the data acquired on the molecule or the surface. The ensuing normalized presented curves represents the repeatable (dI/dV)/(I/V) spectrum we obtained.

**Theoretical methods:**
**Charge transfer parameters and rate constant:**

The basis of our model rests on Fermi's golden rule (FGR), which states[54],

$$P_{j \leftarrow i} = \frac{2\pi}{\hbar} |V_{ij}|^2 \delta(\varepsilon_i - \varepsilon_j) \tag{3}$$

where $\varepsilon_{i/j}$ are the energies of the quantum states involved, and it clearly shows that the integral of the probability (a rate) is nonzero only when $\varepsilon_i$ and $\varepsilon_j$ are degenerate. In other words, the Dirac delta involved in the FGR expression is the function that ensures that when an initial and a final state are degenerate, the transfer probability is largest. In Eq. (1), we slightly relax this condition and we assume that as long as the transfer occurs downhill in energy, it is allowed. That is, we assume that vibrations will be very efficient in dissipating excess energy in both hole and electron transfer process. Hence, the peculiar definition of the $\Theta_{ij}$ in Eq.(2). To realize this model, alongside a sum over all the possible final states, in Eq. (1) of the main text there is also a sum over the initial states. This is simply a mathematical construction (together with the use of another Dirac delta function in the first summation) to find a molecular state in the acceptor molecule that is in resonance with the prepared ionic species at energy E, which is, in a first approximation, the STM tip Fermi energy. We introduced $P_I(\varepsilon_i)$ as a weighting function for a given initial state energy. This mimics the role of the STM tip excitation process. An equivalent partial DOS is not considered for the acceptor molecule because the model assumes that all regions of the acceptor are open to accept a charge modulated by the coupling term $|V_{ij}|^2$. The Fermi-Dirac functions, $f_I$ and $f_F$, are also introduced to make sure that only filled electronic states for hole transfer and empty states for electron transfer are considered while still employing unconstrained quantum state's summations.

The Hamiltonian and overlap matrix elements are obtained by the following single-particle transfer integrals (also known as Fragment Orbital Method[55]) involving HOMO and LUMO orbitals ($\phi^{H/L}$):

$$H_{ij} = \langle \Psi_i | \hat{H}_{el} | \Psi_j \rangle \cong \langle \phi_i^{H/L} | \hat{h}_{KS} | \phi_j^{H/L} \rangle \tag{4}$$

$$S_{ij} = \langle \Psi_i | \Psi_j \rangle \cong \langle \phi_i^{H/L} | \phi_j^{H/L} \rangle \tag{5}$$

Where $\hat{h}_{KS}$ is the single-particle Kohn-Sham Hamiltonian and $\phi_i^{H/L}$ are either the HOMO (hole) or LUMO (electron) orbitals for either donor (*i*) or acceptor (*j*) fragments. The electronic coupling, $V_{ij}$, is generally represented as the Hamiltonian coupling between Löwdin orthogonalized states, taking the form:

$$V_{ij} = \langle \Psi_i | \hat{H}_{el} | \Psi_j \rangle = \frac{1}{1 - S_{ij}^2} \left( H_{ij} - S_{ij} \frac{(H_{ii} + H_{jj})}{2} \right) \tag{6}$$

Above, $H_{ii}$ and $H_{jj}$ are either the HOMO (hole) or LUMO (electron) site energies of donor and acceptor, respectively.

**Density Functional Theory:** All calculations are performed employing the Amsterdam Density Functional (ADF) program[56]. The Hybrid exchange-correlation functional B3LYP[57], which contains approximately 20% of exact exchange, is used along with the TZP basis set of Slater-Type Orbitals. Relativistic effects should play a minor role for first-row transition metal elements (such as Fe), however, it is accounted for employing the scalar ZORA approximation[58]. The procedure to obtain the electronic coupling using the Transfer Integrals (TI) method[59] starts with the evaluation of the molecular orbitals and corresponding energies for each fragment involved in the transfer. The monomers are computed in the neutral ground state, and the dimer is described by simple direct sum of the monomer's density matrices. The energy levels are computed by a single point calculation of the isolated molecule and the PDOS are calculated individually for each pyrrole site by utilizing the DOS program available in the ADF suite of programs.

**Figures captions:**

**Figure 1:** (a) sketch of the charge transfer principle in a homo-molecular dyad. (b) ball and stick representation of the Fe-TPP molecule. The white, light gray, blue and dark gray ball represents the hydrogen, carbon, nitrogen and iron atoms, respectively. (c) (110 x 110 Å²) STM topography ($V_s$ = -2.3 V, I = 1.5 pA) of the insulating layer of $CaF_2$ stripes following the adsorption of the Fe-TPP molecules. (d) and (e) (27.5 x 27.5 Å²) STM topographies ($V_s$ = -2.3 V, I = 1.5 pA) of the Fe-TPP molecule (left) and the same with a superimposed wireframe of the molecule (right) for the CL and CR conformations, respectively.

**Figure 2:** (a) (27.5 x 27.5 Å²) STM topographies ($V_s$ = -2.3 V, I = 1.5 pA) of a Fe-TPP in the CL configuration with the eight studied excitation locations. (b) Variation of the measured quantum yield for the CL → CR movement for the eight defined positions and for the two excitation biases -2.5 and 2.5 volts. (c) (27.5 x 27.5 Å²) STM topographies ($V_s$ = -2.2 V, I = 1.5 pA) of a Fe-TPP in the CR configuration with the eight studied excitation locations. (d) Variation of the measured quantum yield for the CR → CL movement for the eight defined positions and for the two excitation biases -2.5 and 2.5 volts.

**Figure 3:** (a) (27.5 x 55 Å²) STM topographies ($V_s$ = -2.3 V, I = 1.0 pA) of a Fe-TPP dyad in the CL-CL (left) and CL – CR (right) configurations, i.e. before and after a CT. In the left topography, are indicated the four excitation positions (red dots) chosen to study the CT process in the dyad. (b) Variation of the quantum yield of the CT process induced in the dyad for the four defined locations in (a). (c) Detailed (27.5 x 45.5 Å²) STM topography ($V_s$ = -2.2 V, I = 1.0 pA) of the Fe-TPP dyad where the various distances $d_1$ to $d_4$ are indicated. (d) Variations of the CT yield as a function of the four considered distances $d_1$ to $d_4$ as defined in (c) for the two considered biases -2.5 V and 2.5 V. The red dashed curve is the fitting curve of the data for holes (-2.5 V) by the expression $\Gamma(d) = \gamma_0 + \gamma_1 \exp(-d/\beta)$.

**Figure 4:** (a) Ball and stick sketch of the FeTPP molecule after two types of rotation along $\gamma$ or $\theta$ with the white, gray, purple and light blue atoms representing the hydrogen, carbon, nitrogen and iron, respectively. (b) Table of the calculated spatial distribution of the frontier HOMO and LUMO orbitals for three values of $\theta$.

**Figure 5:** (a) Ball and stick representation of the FeTPP dyad used to compute the variation of $\Gamma(E)$ for the two values of $\theta$. The white, gray, purple and light blue atoms represent the hydrogen, carbon, nitrogen and iron, respectively. The red, blue, green and black circles indicate the atoms of the pyrroles groups considered to compute the partial density of state $P_I$. (b) and (c) computed variations of $\Gamma(E)$ for the electron transfer as a function of the energy E-$E_f$ for the four location $p_1$ to $p_4$ and for $\theta = 45°$ and $\theta = 10°$. (d) and (e) computed variation of $\Gamma(E)$ for the hole transfer as a function of the energy E-$E_f$ for the four location $p_1$ to $p_4$ and for the two values of $\theta = 45°$ and $10°$ (f). Log scale variation of computed $\Gamma(E)$ as a function of the distances $d_1$ to $d_4$ as described in Fig. 3c.

**Figure 6:** Energetic diagram of twenty computed orbitals of the FeTPP molecule with $\theta = 10°$ compared with the band diagram structure of the surface. The electrostatic potential $\mu_e$ is located in the middle of the molecular gap HOMO-LUMO, which is itself centered at the Fermi level of the (n-type doped) silicon surface. Relevant energies/biases (-1.8V, -2.5 V) are recalled for clarity. The insert is the variation of the Log($\tau/\tau_{HOMO}$) as a function of the energy E-$\mu_e$ (see text).

**Figure 7:** (a) Energetic diagram of the first step (step 1) of the CT process occurring in the molecular dyad associated with a 2D sketch of the STM tip, the molecular dyad and the surface as positioned during the excitation process in (lower part). Step 1 describes the ionization of the excited pyrrole group (Pyr$^N$) in the donor molecule that leads to an excited cation PES Pyr$^+$. At the energy of Pyr+, the acceptor fragment of the dyad exhibit a double well PES Porph+Aryl$^N$ in the neutral state that can accept the hole of the donor fragment.  (b) Step 2 of the CT describing the neutralization of the excited pyrrole getting an excess of kinetic energy *ke* while the acceptor fragment is ionized and thus reaches a PES at higher energy Porph+Aryl$^+$. The ensuing sketch below this panel indicates that the STM tip is still exciting the donor molecule during this very short time while the charge image of the cation is displaced underneath the acceptor molecule. (c) Step 3 of the CT process in the Fe-TPP dyad involving the neutralization of the second fragment of the dimer (acceptor) leading to the vibrational excitation of the second fragment via *ke'*. The vibrational population spread over the molecule to involve one aryl group which is responsible of the final acceptor movement (see sketch below panel (c)). The neutralization occurs via surface mediated charge transfer. The life times of each process arise from experimental and theoretical estimations (see text).

**References:**

[1]Bergfiled, J. P.; Ratner, M. A. Forty years of molecular electronics: Non-equilibrium heat and charge transport at the nanoscale. *Phys. Stat. Sol. B* **2013**, 250, 2249.

[2]Huynh, W. U.; Dittmer J. J.; Alivisatos A. P. Hybrid Nanorod-Polymer Solar Cells. *Science* **2002**, 295, 2425.

[3] Wohlgenannt, M.; Jiang, X. M.; Vardeny, Z. V.; Janssen, R. A. J. Conjugation-Length Dependence of Spin-Dependent Exciton Formation Rates in Π-Conjugated Oligomers and Polymers. *Phys. Rev. Lett.* **2002**, 88, 197401.

[4] Jäckel, F.; Perera, U. G. E.; Iancu, V.; Braun, K.-F.; Koch, N.; Rabe, J. P.; Hla, S.-W. Investigating Molecular Charge Transfer Complexes with a Low Temperature Scanning Tunneling Microscope. *Phys. Rev. Lett.* **2008**, 100, 126102.

[5] Porath, D.; Bezryadin, A.; de Vries, S.; Dekker C. Direct measurement of electrical transport through DNA molecules. *Nature* **2000**, 403, 635.

[6] Dwayne Miller, R.J. Time scale issues for charge transfer and energy storage using semiconductor junctions. Solar energy Materials and Solar Cells, **1995**, 38, 331.

[7] Barbara, P.F; Meyer, T.J., Ratner, M.A. Contemporary Issues in electron transfer research. J. Phys. Chem. **1996**, 100, 13178.

[8] Novoderezhkin, V.I.; Romerob, E.; van Grondelle, R. How exciton-vibrational coherences control charge separation in the photosystem II reaction center. PCCP, **2015**, 17, 30828.

[9] Zobel, J.P.; Nogueira, J.J. González, L. Quenching of Charge Transfer in Nitrobenzene Induced by Vibrational Motion. J. Phys. Chem. Lett. **2015**, 6, 3006.

[10] Delor, M.; Keane, T.; Scattergood, P.A.; Sazanovich, I.V. Greetham, G.M.; Towrie, M.; Meijer, A.J.H.M. Weinstein, J.A. On the mechanism of vibrational control of light-induced charge transfer in donor–bridge–acceptor assemblies. Nat. Chem. **2015**, 7, 689.

[11] Duchemin, I. ; Blase, X. ; Resonant hot charge-transfer excitations in fullerene-porphyrin complexes: Many-body Bethe-Salpeter study. Phys. Rev. B, **2013**, 87, 245412.

[12] Small organic molecules on surfaces, Sitter, H.; Draxl, C.; Ramsey, M. (Eds.), Springer series in material science, ISBN 978-3-642-33848-9.

[13] Cheung, D.L. ; Troisi, A. Modelling charge transport in organic semiconductors: from quantum dynamics to soft matter, Phys. Chem. Chem. Phys., **2008**, 10, 5941.

[14] Neaton, J.B.; Hybersten, M.S.; Louie, S.G.; Renormalization of Molecular Electronic Levels at Metal-Molecule Interfaces. Phys. Rev. Lett. **2006**, 97, 216405.

[15] Piva, P.G.; DiLabio, G.A. Pitters, J.L. Janik Zikovsky, Moh'd Rezeq, Stanislav Dogel, Werner A. Hofer & Robert A. Wolkow, Field regulation of single-molecule conductivity by a charged surface atom, Nature **2005**, 435, 658.

[16] Gating of single molecule junction conductance by charge transfer complex formation, Nanoscale, **2015**, 7, 18849-18955.

[17] Labidi, H.; Pinto, P.; Leszczynski, L.; Riedel D. Exploiting a single intramolecular conformational switch to probe charge transfer dynamics at the nanoscale. Phys. Chem. Chem Phys. **2017**, 19, 28982.

[18]Chiaravalloti, F.; Dujardin, G.; Riedel, D. Atomic scale control of hexaphenyl molecules manipulation along functionalized ultra-thin insulating layer on the Si(100) surface at low temperature (9 K). *J. Phys.: Condens. Matter.* **2015**, 27, 054006.

[19]Lastapis, M.; Martin, M.; Riedel, D.; Hellner, L. et al. Picometer-scale electronic control of molecular dynamics inside a single molecule, *Science* **2005**, 308, 1000.

[20]Bellec, A.; Chaput, L.; Dujardin, G.; Riedel, D.; Stauffer, L.; Sonnet, P. Reversible charge storage in a single silicon atom. Phys. Rev. B, **2013**, 88, 241406.

[21]Yengui, M.; Duverger, E.; Sonnet, P. Riedel, D. A two dimensional On/Off switching device based on anisotropic interactions of atomic quantum dots on Si(100):H. Nature Communication, **2017**, 8, 2211.

[22]Imada, H. ; Miwa, K.; Imada, M.I., Kawahara, S. Real-space investigation of energy transfer in heterogenous molecular dimers, Nature **2016**, 538, 364 – 368.

[23]Zhang, Y.; Luo, L.; Zhang, Y.; Yu, Y.J.; Kuang, Y.M.; Zhang, L.; Meng, Q. S. Luo, Y.; Yang, J.L.; Dong Z.C.; Hou, J.G. Visualizing coherent intermolecular dipole–dipole coupling in real space. Nature, **2016**, 531, 623-627.

[24]Pavanello, M.; Neugebauer, J. Linking the historical and chemical definitions of diabatic states for charge and excitation energy transfer reactions in condensed phase. J. Chem. Phys. **2011**, 135, 134113.

[25]Subotnik, J.; Cave, R.J.; Steele, R.P.; Shenvi, N. The initial and final states of electron and energy transfer processes: Diabatization as motivated by system-solvent interactions. J. Chem. Phys. **2009**,130, 234102.

[26]Mattay, J. Charge transfer and radical ions in photochemistry. *Angew. Chem. Int. Ed. Engl*, **1987**, 26, 825-845.

[27]Novoderezhkin, V.I.; Romero, E.; Grondelle, R.V. How exciton-vibrational coherences control charge separation in the photosystem II reaction center. Phys.Chem.Chem.Phys. **2015**, 17, 30828.

[28]Ruggieri, C.; Rangan, S.; Bartynski, R.A., Galoppini, E., Zinc(II) Tetraphenylporphyrin on Ag(100) and Ag(111): Multilayer Desorption and Dehydrogenation, *J. Phys. Chem. C*, **2016**, 120, 7575

[29]Furmansky, J.; Sasson, H.; Liddell, P.; Gust, D.; Ashkenasydb, N.; Visoly-Fisher, I. Porphyrins as ITO photosensitizers: substituents control photo-induced electron transfer direction. J. Mater. Chem., **2012**, 22, 20334.

[30]Reuter, M.G.; Boffi, N.M.; Ratner, M.A. Seideman, T. The role of dimensionality in the decay of surface effects. J. Chem. Phys. **2013**, 138, 084707.

[31] Potapenko, D. V. ; Li, Z. ; Osgood, R.M.; Osgood, R.M. Dissociation of Single 2-Chloroanthracene Molecules by STM-Tip Electron Injection, J.Phys. Chem. C, **2012**, 116, 4679.

[32] P.; Liljeroth, J.; Repp, G.; Meyer, Current-induced hydrogen tautomerization and conductance switching of naphthalocyanine molecules, Science, **2007**, 317, 1203.

[33] A. ,Yu; S., Li ; G. Czap; W., Ho Tunneling-Electron-Induced Light Emission from Single Gold Nanoclusters, NanoLett. **2016**, 16, 5433.

[34] Electron (or hole) transfer process can be induced in both ways where the donor transfer its electron (or hole) to the acceptor depending on the initial formation of an anion or cation in the donor molecule.

[35] Skourtis, S.; Effects of initial state preparation on the distance dependence of electron transfer through molecular bridges and wires. J. Chem. Phys. **2003**, 119, 6271.

[36] Liao, M.S.; Scheiner, S. Electronic structure and bonding in metal porphyrins, metal=Fe, Co, Ni, Cu, Zn. *J. chem. Phys.* **2002**, 117, 205.

[37] Several other molecular conformations are also observed either naturally after the adsorption or via STM manipulation. These other conformations are not shown and exploited in this work.

[38] Labidi, H.; Kantorovitch, L.; Riedel, D. Atomic-scale control of hydrogen bonding on a bare Si(100)-2×1 surface. *Phys. Rev. B* **2012**, 86, 165441.

[39] Riedel, D. Single molecule manipulation at low temperature and laser scanning tunnelling photo-induced processes analysis through time resolved dynamics studies, *J. Phys.: Condens. Matter.,* **2010**, 22, 264009.

[40] Bellec, A; Riedel, D.; Dujardin, G.; Boudrioua, O.; Chaput, L.; Stauffer, L.: Sonnet, P.; Nonlocal Activation of a Bistable Atom Through Surface State Charge Transfer on Si(100)-2x1:H. *Phys. Rev. Lett.*, **2010**, 105, 048302.

[41] Riedel, D.; Cranney, M.; Martin, M.; et al. Surface-Isomerization Dynamics of trans-Stilbene Molecules Adsorbed on Si(100)-2x1. *J. Am. Chem. Soc.*, **131**, 5414 (2009).

[42] Labidi, H.; Sonnet, P.; Riedel, D.; Electronic Control of the Tip-Induced Hopping of an Hexaphenyl-Benzene Molecule Physisorbed on a Bare Si(100) Surface at 9 K. *J. Phys. Chem. C* **2013**, 117, 13663.

[43] The chemical Physics of solid surfaces, Woodruff D.P. (Eds.) Surface Dynamics-Academic Press, Elsevier (2007). ISBN: 978-0-444-52756-1.

[44] Shiatory, A. Shiotari A. (2017) Symmetry Correlation between Molecular Vibrations and Valence Orbitals: NO/Cu(110) and NO/Cu(001). In: Reactivity of Nitric Oxide on Copper Surfaces. Springer Theses (Recognizing Outstanding Ph.D. Research). Springer, Singapore.

[45] High, J. S.; Rego, L.G.C.; Jakubikova, E. Quantum Dynamics Simulations of Excited State Energy Transfer in a Zinc–Free-Base Porphyrin Dyad. J. Phys. Chem. A, **2016**, 120, 8075-8084.

[46] Tsuchiya, T.; Jakubikova, E. Role of noncoplanar conformation in facilitating ground state hole transfer in oxidized porphyrin dyads. J. Phys. Chem. A, **2012**, 116, 10107.

[47] Shunsuke, S.; Drummen, G.P.C.; Konishi, G.I. Recent advances in twisted intramolecular charge transfer (TICT) fluorescence and related phenomena in materials chemistry, J.Mater.Chem.C, **2016**, 4, 2731.

[48] Hernandez-Fernandez, F.; Pavanello, M.; Visscher, L. Effect of metallation, substituents and inter/intra-molecular polarization on electronic couplings for hole transport in stacked porphyrin dyads. Phys. Chem. Chem. Phys., **2016**, 18, 21122-21132.

[49] Demchenko, A.P. ; Tomin, V.I. ; Chou, P.T. Breaking the Kasha rule for more efficient photochemistry , Chem. Rev. **2017**, 117, 13353.

[50] K., Senthilkumar ; F.C. Grozema ; C.F. Guerra ; F.M. Bickelhaupt; F.D., Lewis et al. Absolute rates of hole transfer in DNA, J. Am. Chem. Soc. **2005**, 127, 14894.

[51] McConnel H. F. Intramolecular Charge transfer in aromatic free radicals, *J. Chem. Phys.* **1961**, 35, 508-515.

[52] Giese, B.; Wessely, S.; Spormann, M.; Lindemann, U.; Meggers, E. Michel-Beyerle, M. E. On the Mechanism of Long-Range Electron Transfer through DNA. *Angew. Chem Int. Ed*. **1999**, 38, 996.

[53] Riedel, D; Bocquet, M.L. ; Lesnard, H. et al. Selective Scanning Tunnelling Microscope Electron-induced Reactions of Single Biphenyl Molecules on a Si(100) Surface. J. Am. Chem. Soc., **2009**, 131, 7344.

[54] Chemical Dynamics in Condensed Phases: Relaxation, Transfer, and Reactions in Condensed Molecular Systems, Nitzan, A., Oxford University press, 2006.

[55] Oberhofer, H.; Blumberger, J. Revisiting electronic couplings and incoherent hopping models for electron transport in crystalline C60 at ambient temperatures. Physical Chemistry Chemical Physics, **2012**, 14, 13846.

[56] ADF2017, SCM, Theoretical Chemistry, Vrije Universiteit, Amsterdam, The Netherlands, http://www.scm.com.

[57] Lenthe, E.V.; Baerends, E.J.; Snijders, J.G.; Relativistic regular two-component Hamiltonians. J. Chem. Phys. **1993**, 99, 4597.

[58] Lenthe,E.V.; Baerends, E.J.; Snijders, J.G.; Relativistic total energy using regular approximations. J. Chem. Phys. **1994**, 101, 9783.

[59] Senthilkumar, K.; Grozema, F. C.; Bickelhaupt, F. M.; Siebbeles, L. D. A. Charge transport in columnar stacked triphenylenes: Effects of conformational fluctuations on charge transfer integrals and site energies. J. Chem. Phys. **2003**, 119, 9809.

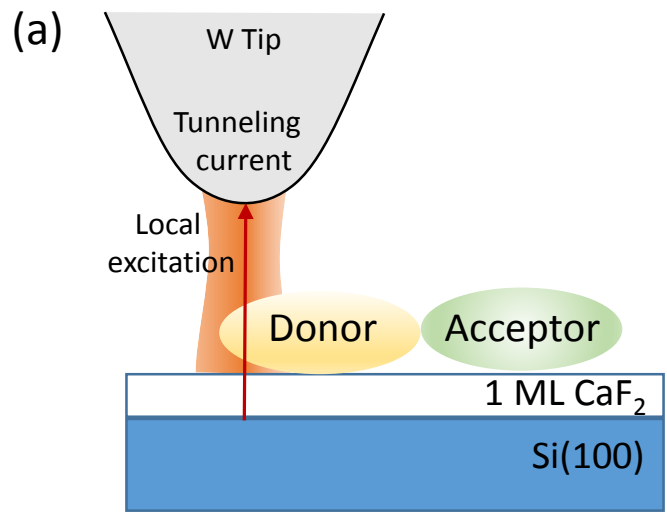
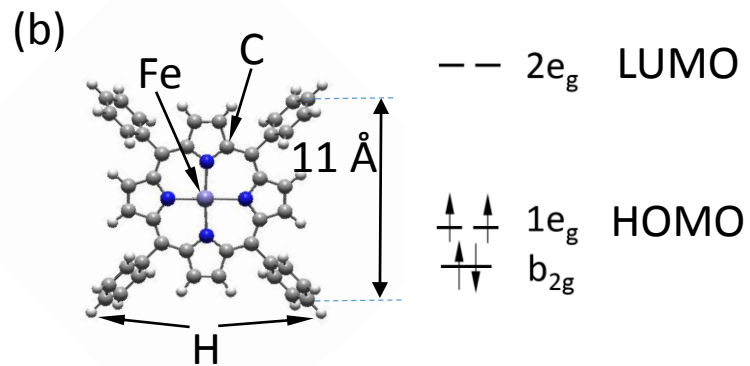
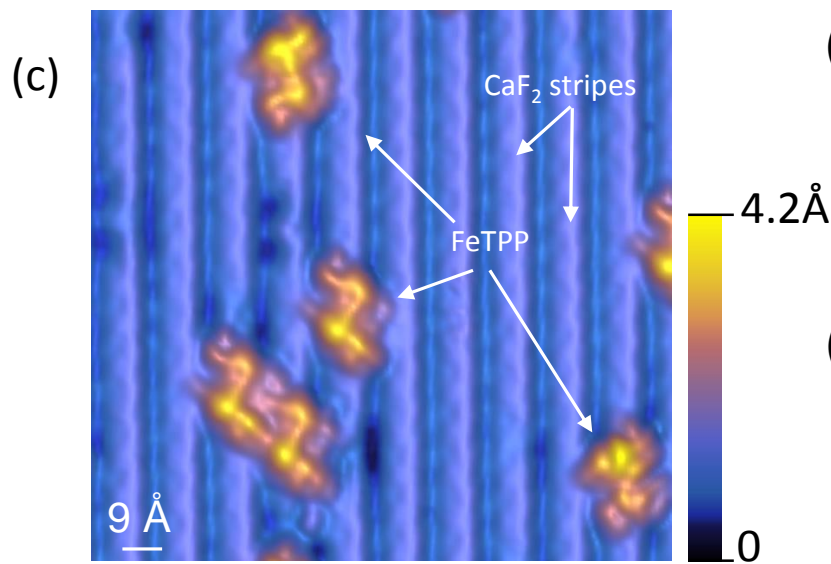
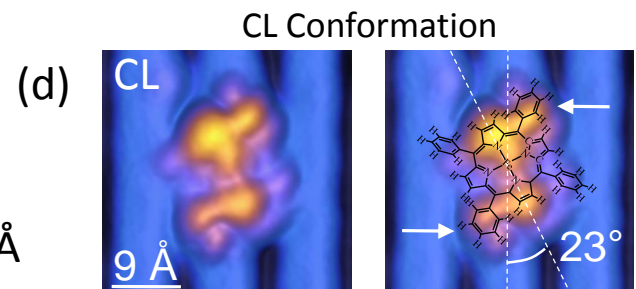
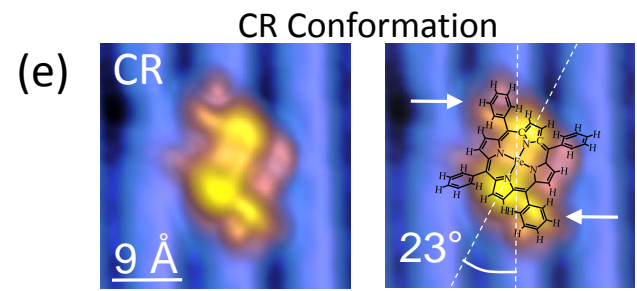

Ramos et al. Figure 1

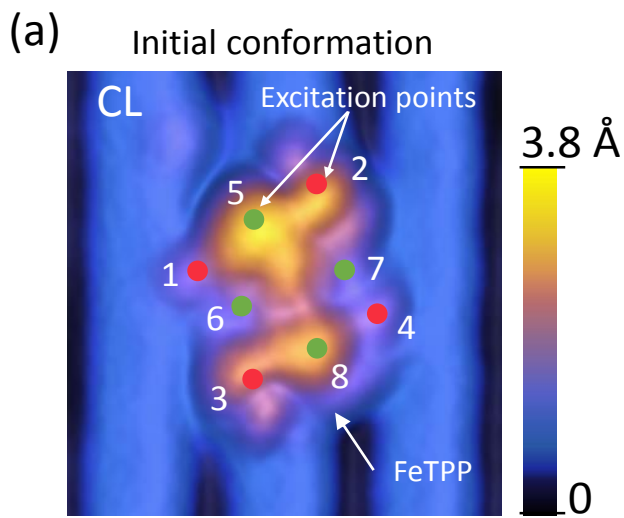
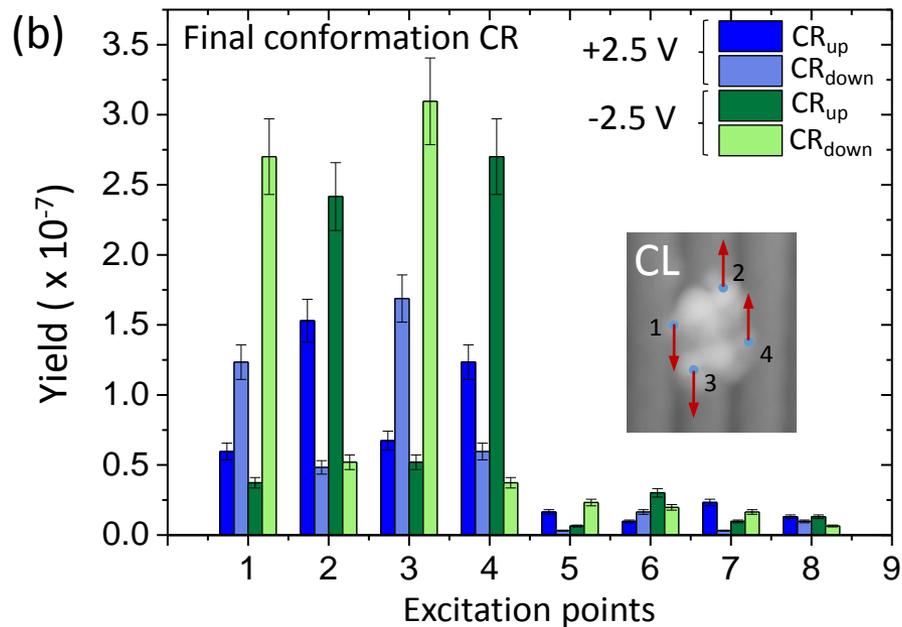
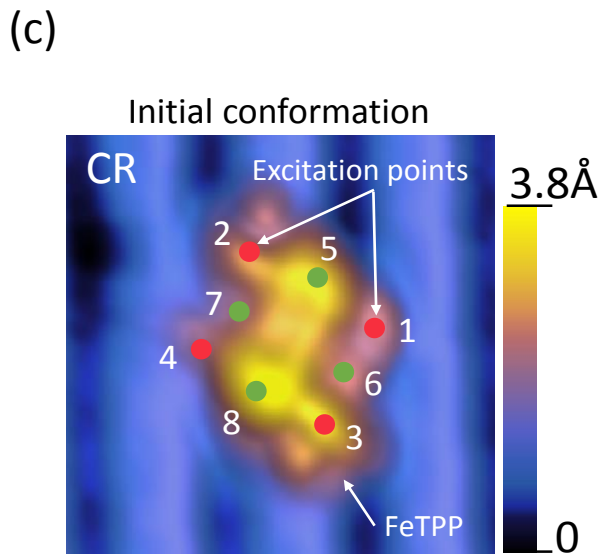
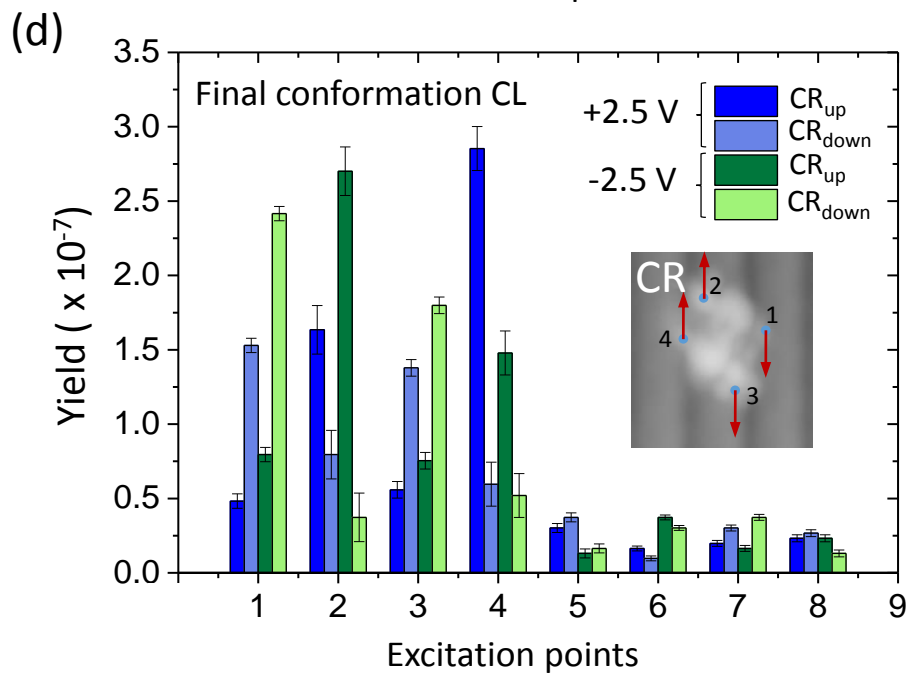

Ramos et al. Figure 2

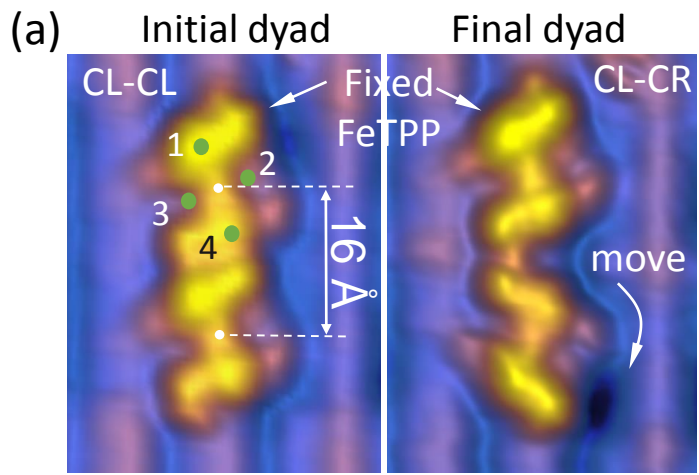
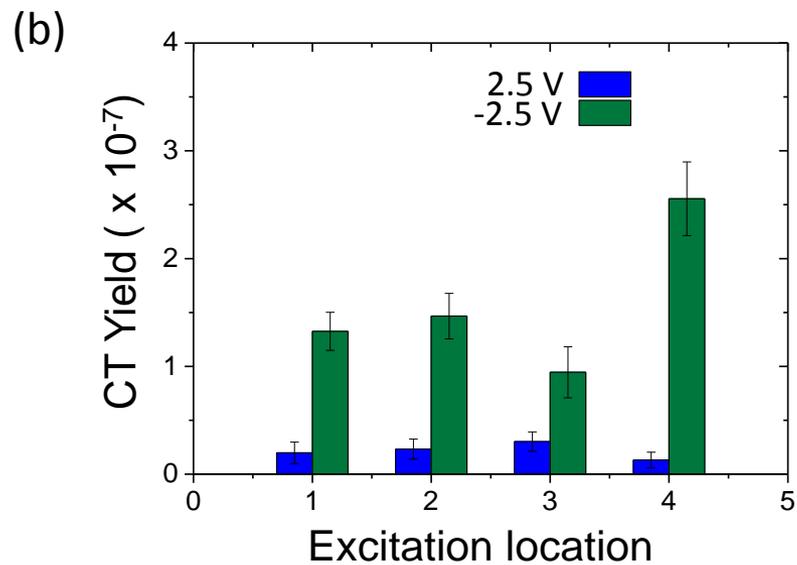
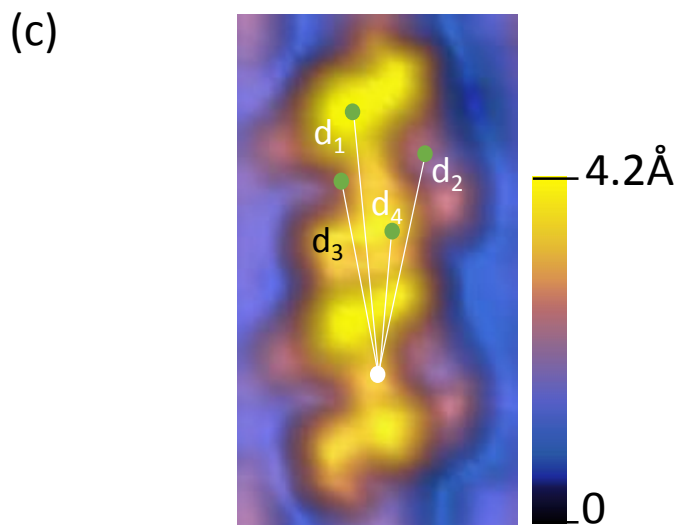
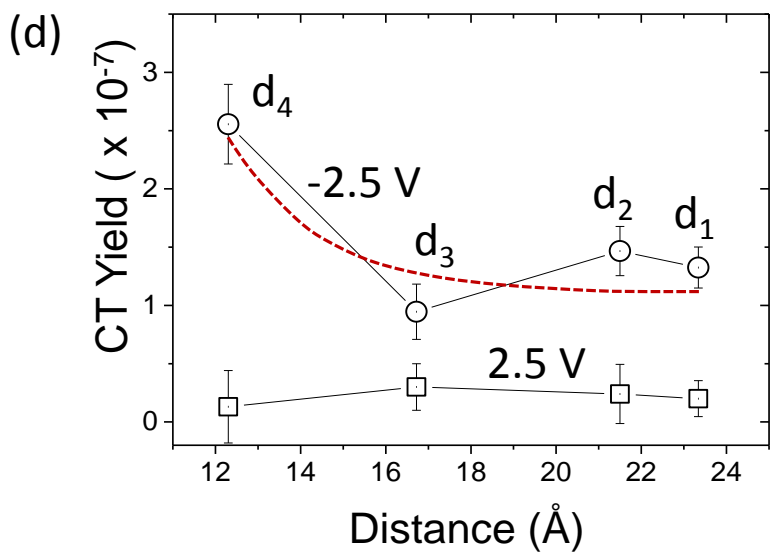

Ramos et al. Figure 3

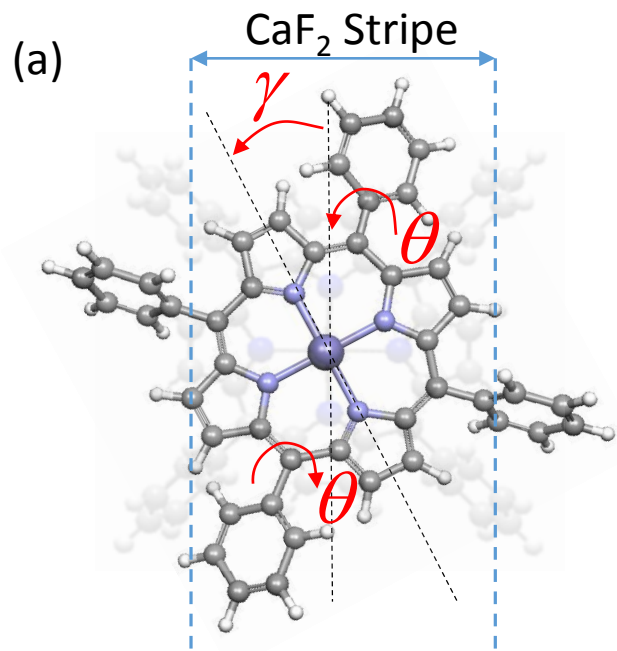

Ramos et al. Figure 4

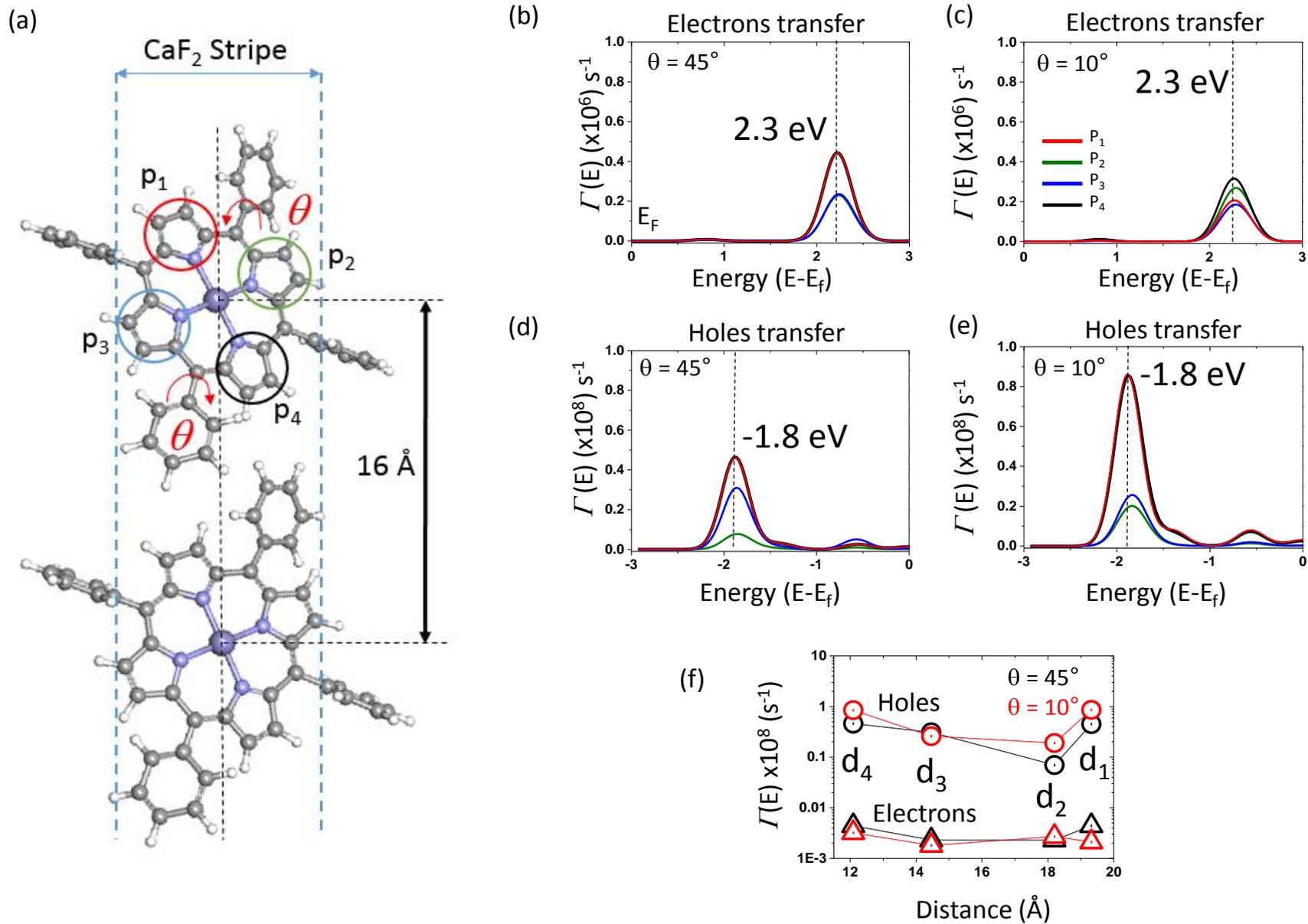

Ramos et al. Figure 5

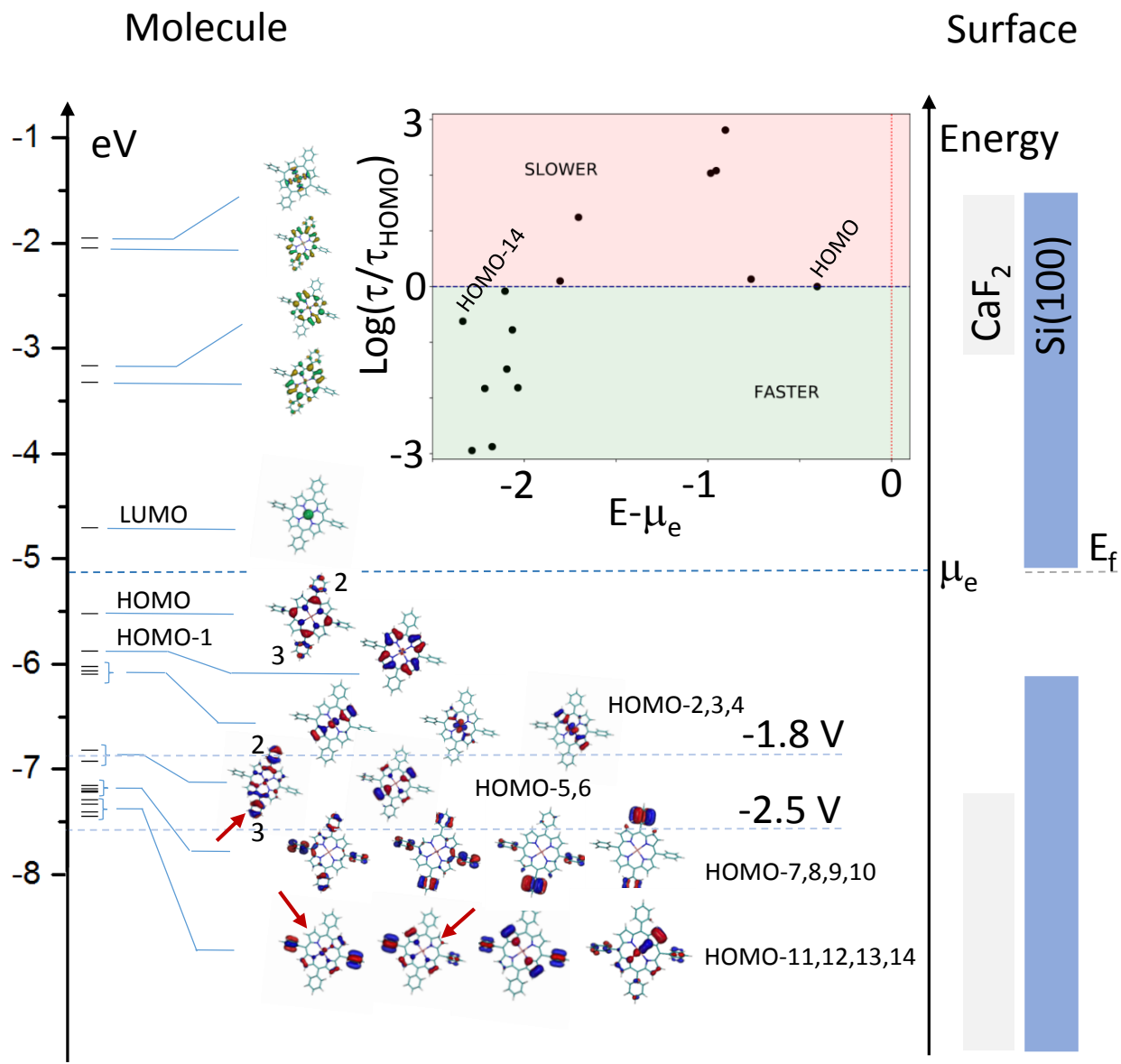

Ramos et al. Figure 6

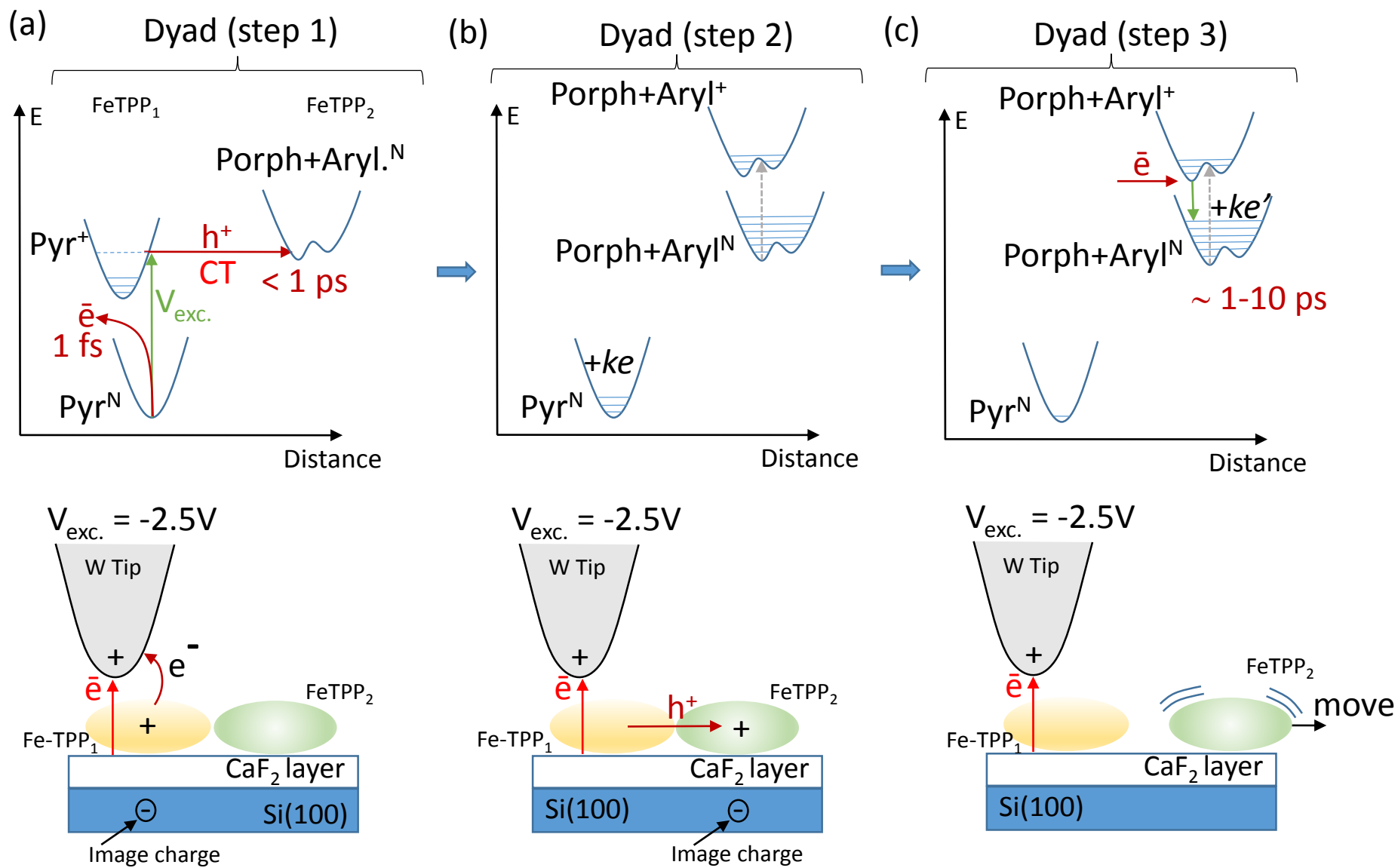

Ramos et al. Figure 7